\documentclass[sigconf]{acmart} 
\AtBeginDocument{%
  \providecommand\BibTeX{{%
    \normalfont B\kern-0.5em{\scshape i\kern-0.25em b}\kern-0.8em\TeX}}}

\copyrightyear{2025}
\acmYear{2025}
\setcopyright{acmlicensed}\acmConference[IMX '25]{ACM International Conference on Interactive Media Experiences}{June 3--6, 2025}{Niterói, Brazil}
\acmBooktitle{ACM International Conference on Interactive Media Experiences (IMX '25), June 3--6, 2025, Niterói, Brazil}
\acmDOI{10.1145/3706370.3727866}
\acmISBN{979-8-4007-1391-0/2025/06}

\usepackage{CJKutf8}
\begin{document}
\begin{CJK*}{UTF8}{min}
\title[Entertainers Between Real and Virtual]{Entertainers Between Real and Virtual --- Investigating Viewer Interaction, Engagement, and Relationships with Avatarized Virtual Livestreamers}

\author{Michael Yin}
\affiliation{
  \institution{University of British Columbia}
  \city{Vancouver}
  \state{BC}
  \country{Canada}
  \postcode{V6T 1Z4}
  \orcid{0000-0003-1164-5229}
}
\email{jiyin@cs.ubc.ca}

\author{Chenxinran Shen}
\affiliation{
  \institution{University of British Columbia}
  \city{Vancouver}
  \state{BC}
  \country{Canada}
  \postcode{V6T 1Z4}
  \orcid{0009-0009-9967-8017}
}
\email{elise.shen007@gmail.com}

\author{Robert Xiao}
\affiliation{
  \institution{University of British Columbia}
  \city{Vancouver}
  \state{BC}
  \country{Canada}
  \postcode{V6T 1Z4}
  \orcid{0000-0003-4306-8825}
}
\email{brx@cs.ubc.ca}

\begin{abstract}
Virtual YouTubers (VTubers) are avatar-based livestreamers that are voiced and played by human actors. VTubers have been popular in East Asia for years and have more recently seen widespread international growth. Despite their emergent popularity, research has been scarce into the interactions and relationships that exist between avatarized VTubers and their viewers, particularly in contrast to non-avatarized streamers. To address this gap, we performed in-depth interviews with self-reported VTuber viewers (n=21). Our findings first reveal that the avatarized nature of VTubers fosters new forms of theatrical engagement, as factors of the virtual blend with the real to create a mixture of fantasy and realism in possible livestream interactions. Avatarization furthermore results in a unique audience perception regarding the identity of VTubers --- an identity which comprises a dynamic, distinct mix of the real human (the voice actor/actress) and the virtual character. Our findings suggest that each of these dual identities both individually and symbiotically affect viewer interactions and relationships with VTubers. Whereas the performer's identity mediates social factors such as intimacy, relatability, and authenticity, the virtual character's identity offers feelings of escapism, novelty in interactions, and a sense of continuity beyond the livestream. We situate our findings within existing livestreaming literature to highlight how avatarization drives unique, character-based interactions as well as reshapes the motivations and relationships that viewers form with livestreamers. Finally, we provide suggestions and recommendations for areas of future exploration to address the challenges involved in present livestreamed avatarized entertainment. 
\end{abstract}

\ccsdesc[500]{Human-centered computing~Empirical studies in HCI}

\keywords{VTuber; livestreaming; virtual interactions; virtual avatars}


\maketitle

\section{Introduction}

Livestreaming (or simply streaming) has become a massively popular media phenomenon and a reported billion-dollar industry \cite{livestreamrevenue,wang2018research,johnson2019impacts,lu2018watch}. Day after day, streamers record and publicize real-time videos that can be watched by millions of viewers around the world on streaming platforms, such as Twitch and YouTube Live \cite{li2020systematic}. These platforms allow streamers to distribute media content to their audience \cite{pires2015YouTube, anderson2018getting}. The importance of livestreaming as a modern media phenomenon has resulted in numerous academic studies on understanding the various unique facets that affect both viewer and streamer motivations and interactions. For example, past research has considered how factors such as social interaction or a sense of community drive viewership \cite{sjoblom2017people, hilvert2018social}. Livestreaming also creates different forms of real-time interactions compared to traditional entertainment media, with comparisons to performance and theatre \cite{li2019live, scully2017playing, liud2021virtual}. 

Past research has primarily focused on real-person streamers --- streamers that show their real face in front of the camera. Realness and authenticity have been discussed as integral parts of the relationship between streamer and viewer \cite{lu2018watch}. However, the phenomenon of Virtual YouTubers (VTubers) --- virtual live-streaming avatars voiced by humans --- has recently carved a niche within the livestreaming space. Compared to traditional forms of livestreaming, VTubers use an \textbf{avatar} that mimics the expressions and movements of a ``Nakanohito'' or ``performer'' --- the voice actor/actress that brings the avatar to life \cite{lu2021kawaii}. Thus, rather than seeing a human, audience members see an avatarized character perform. VTubers have seen popularity within East Asian countries since their emergence in 2016, and nowadays even intersect with mainstream media \cite{mainstream1, mainstream2}. However, their global widespread rise is a more recent event, hastened by new generations of VTubers speaking a wide variety of different languages. Nowadays, VTuber corporations --- companies that manage and support VTuber talents --- partake in a multi-million dollar global industry \cite{hololiverev}. 

The emergence of VTubers as a social, cultural, and technological phenomenon offers potential insights into audience interactions with virtual, avatarized entertainers played by humans, but could also hark towards an evolving entertainment landscape that blurs the line between the physical and the virtual, between realism and fiction. Previous research work has looked into the interactions and engagements between viewers and VTubers from a human-motivational perspective \cite{lu2021kawaii}. Our study extends this work, focusing on the specific idiosyncratic element of \textbf{avatarization} during livestreaming, and how the distinct lack of a front-facing real-person affects the interactions, motivations, and relationships formed between the viewer and streamer. We incorporate past sociological and psychological viewpoints regarding both streaming and avatar-based interactions and we look into how the avatarization aspect of VTubers shapes both the scope of audience-facing interactions and the audience's underlying intrinsic motivations and experiences. Thus, we structure this work as an exploration of the following research questions:

\begin{itemize}
    \item \textbf{RQ1}: How does avatarization foster novel interactions in livestreaming?
    \item \textbf{RQ2}: How are viewer motivations and streamer-viewer relationships affected by avatarization? 
    \item \textbf{RQ3}: What challenges and limitations arise in livestreaming due to avatarization, and how could such challenges be addressed?
\end{itemize}

We performed interviews with regular viewers of VTubers, ultimately finding that \textbf{avatarization of livestreams offer unique affordances and challenges} --- VTubers can perform in ways that real-person livestreamers cannot and vice-versa. Being an interactive entertainer who is not fully real and not fully virtual also creates unique viewer perspectives and motivations for watching, as viewers engage with and form relationships with a performer that blends separate identities. We highlight \textbf{how factors of avatarization mediate the relationship between a streamer's audience and the streamer themselves} and discuss \textbf{how such factors also affect identity, audience engagement, and so forth}. We believe that the blending of identities facilitated by avatarization warrants further exploration, as it offers a unique opportunity to understand how identity shifts can be effectively managed to enhance experiences in computer-mediated communication research.

\section{Background and Related Works}

To contextualize our study, we first consider the history and state of the VTuber industry. We then consider academic research into virtual performance and engagement, looking at methods of interaction that arise with avatarized agents. Lastly, we explore the wealth of literature on livestreaming, considering the perceptions, motivations, and relationships formed from this practice. 

\subsection{Background and Prior Research on VTubers}

The rise in accessible live media expression through livestreaming, combined with advancements in modelling and tracking, has led to the development of VTubers --- entertainers that deliver virtual video content behind an animated virtual avatar \cite{lu2021kawaii}. Although the audience of such content sees an animated virtual character, they hear the voice of a real person --- the performer that plays the character \cite{lu2021kawaii}. The first channel that propelled VTubing into worldwide recognition was Kizuna Ai, an artificial intelligence anthropomorphized as a brunette animated girl \cite{AIchannel}. Debuting in late 2016, Kizuna Ai’s early YouTube videos involved reacting to content, playing games, and just chatting; often featuring the unique incorporation of technological elements such as virtual reality --- e.g. having her, a virtual character, use VR. From then on, VTubing proliferated in popularity \cite{lu2021kawaii}. The continuous development of technology in terms of real-time motion-capture and mapping \cite{realtimemotion} (e.g. using the Live2D software \cite{Live2DCubism}) along with VTuber software suites \cite{akihiko2019reality}, have provided increased accessibility for performers to become VTubers. 

During this time, VTuber corporations started to gain traction. Companies such as Hololive \cite{hololive} and Nijisanji \cite{nijisanji} are for-profit audition-based talent agencies that provide both initial branding as well as a staff and support system to aspiring VTubers. Popular VTubers dominate YouTube Live’s donation message revenue system, with many having pulled millions of dollars from these donations \cite{superchatrev}. Outside of the traditional visual content from streams and videos, VTubers (being public celebrities) may perform in other ways, such as through releasing music albums \cite{unalive} or emceeing events \cite{pixiv}. Additionally, VTubers have also started to move outside of their initial associated subculture and entered the mainstream, for example, as tourism ambassadors \cite{ambassador} and interview hosts with movie stars \cite{mainstream1, mainstream2}. They have also been proposed for virtual mascots at large corporations \cite{crollhime, li2021alime, netflix}. 

Despite the recent global adoption of VTubers, research is relatively scarce --- we highlight the most salient works here. Liudmila considered the design of identity for VTubers, highlighting the challenges presented by VTubers in developing and managing their character, a challenge unique due to the asset-based nature of digital avatars \cite{liudmila2020designing}. Chen et al. studied how VTubers modified their voice features to better align with their designed identity \cite{chen2024conan}. On the more experiential side, Nordvall looked at the online community space within a subset of VTubers, highlighting the materiality and immateriality of space pertaining to digital-physical relationships \cite{nordvall2021down}. Tan studied the level of parasocial attachment to VTubers and its relation to stress relief \cite{tan2023more}, and we explore the aspect of parasocial relationships with avatarized VTubers in this work. However, most comprehensively, Lu et al. performed an interview study with VTuber viewers, considering motivations and relationships formed between audience and entertainer and contrasting these against real-person livestreamers \cite{lu2021kawaii}. 

Our research extends upon this latter study, looking deeply at the specific facets of interaction and engagement that exist specifically due to the \emph{avatarization} aspect of VTubers. Wan and Lu find that VTubers use their avatars to blend real and virtual identities, potentially overacting in performance and increasing inclusiveness in interactions \cite{wanInvestigatingVTubingReconstruction2024}. We tie into their findings from a viewer's perspective and compare areas where such perceptions of identity may potentially diverge. All in all, we go beyond the work of these prior VTuber studies by looking deeply into understanding how the unique characteristic of VTubers intersects \textbf{viewer motivation and relationship-forming with the streamer}, tying those into aspects of sociological research into the livestreaming experience.

\subsection{Avatarized Interactions}

Research into virtual avatars has often been specific to games, e.g., considering the relationship a user feels with their avatars in games, especially in regards to the key concepts of embodiment \cite{hamilton2009identifying, embodiedspace, pasfield2015lost} (the degree in which the user and avatar share a common identity) and presence \cite{schultze2009avatar, parasocialsecondlife} (the sense of being within the virtual world). However, studies have also looked at user perception towards other avatars. Bailey and Blackmore studied how human response and emotional perception may vary based on the gender and appearance of the avatar \cite{bailey2017gender}. Li et al. find that for video game players, avatar attractiveness and appearance agreeability affect perceived friendliness and gamer loyalty \cite{li2018avatar}. Praetorius et al. show that the user’s choice of an avatar may depend on contextual factors such as the social environment in which it is being used \cite{praetorius2021user}. More recently, the rise of virtual reality has allowed for more complex avatarization options, expanding the breadth of research in this area. Rivu et al. studied the effect of avatar gender on the interpersonal distance between friends and strangers \cite{rivu2021when}. These days, improvements in technology have even allowed for high-fidelity, real-time motion-captured avatarization in mixed-reality telepresence \cite{chu2020expressive, Ma_2021_CVPR}. 

Virtual reality avatars can be seen akin to the ones from VTuber contexts because they mimic the motion and expressiveness of a real human; raising potentially similar questions about interactions with others, body ownership, etc. Within this domain, researchers have looked at how features such as avatar representation, facial expressions, motion controls, and so forth, can impact co-presence, trust-building, and behavioural independence \cite{heidickerInfluenceAvatarAppearance2017, kyrlitsiasSocialInteractionAgents2022, luoEffectAvatarFacial2023}. Social VR avatars also offer a way for people to manipulate, craft, and personalize their ``second self'', and can feel like an extension of one's own self \cite{freemanMyBodyMy2020a}. In such a way, harassment toward one's avatar can sometimes equate to harassment toward one's self, and Schulenberg et al. outline the complex and nuanced relationship between avatar and self in this regard \cite{schulenbergCreepyMyAvatar2023}. The Proteus effect in VR outlines some parts of this nuance --- outlining how a user's behaviour and attitude can be affected by their avatar, once again blurring the line between avatar and self-identity \cite{oyanagiPossibilityInducingProteus2022, freemanBodyAvatarMe2021a}. 

We also considered research in the area of avatarized virtual celebrities and influencers --- considering fan engagement and connection with these virtual personas. For example, Miku Hatsune is an avatar-based virtual idol released in 2007 as an anthropomorphism mascot of the Vocaloid software \cite{lu2021kawaii}. She quickly became a worldwide sensation, performing in concerts and starring in games \cite{their2016hatsune, greenwood2013spectral}. Guga found that Miku's virtuality and lack of a singular identity promoted collaborative representation, resulting in a malleable ``aura’’ \cite{Guga2015Hatsune}. Another such virtual celebrity is Lil Miquela, a virtual social media influencer made using CGI \cite{jenna2020celebrity}. Studies have looked at the rise of such virtual celebrities, their relationship to authenticity, and their influence on the habits of followers \cite{hubble2018miquela, jenna2020celebrity, choudhry2022ifelt}. For instance, Sakuma et al. compared the persuasiveness between human and avatarized influencers in terms of willingness to purchase promoted products \cite{sakumaYouTubersVsVTubers2023} --- virtual avatars could be more persuasive in specific domains despite a broader trend toward human persuasiveness. 

Studies have extended on how motivations for following virtual influencers are affected by factors such as information, entertainment, shared passions, social interactions, and so on \cite{louAuthenticallyFakeHow2023, ghzaielWhenArtificialRevolutionizes2025}. Nukhu et al. highlight how people tended to anthropomorphize virtual influencers, applying social norms in their perceptions \cite{nukhuUsersAnthropomorphizeAIBased}. Yet, Ghzaiel et al. \cite{ghzaielWhenArtificialRevolutionizes2025} and Lou et al. \cite{louAuthenticallyFakeHow2023} highlight how challenges still exist over social factors such as perceived authenticity, emotional connection, and weakened parasocial effects; these social and emotional factors can be tied to what Jin and Viswanathan consider the ``need to belong'' as a mediator of using AI-based virtual influences \cite{jinThreatenedEmptySelves2025}. We extend these works to the domain of livestreaming and its emergent relationships between streamer and audience. This is important to study as avatars are increasingly being applied and investigated in group domains such as education \cite{petrakouInteractingAvatarsVirtual2010, parmarHowImmersionSelfAvatars2023}, healthcare \cite{rheuEnhancingHealthyBehaviors2020a, zahediMyRealAvatar2022}, and remote collaboration \cite{piumsomboonMiniMeAdaptiveAvatar2018,  wuUsingFullyExpressive2021}. The integration of AI generation in avatars has seen increased attention as well across such fields \cite{vallisStudentPerceptionsAIGenerated2024, lohRevolutionisingPatientCare}. Vallis and Britton noted how, in an educational environment, students wanted greater interaction with a blended use of human and nonhuman elements \cite{vallisScriptScreenEmergent2024}; though in a different domain, our work highlights how human and nonhuman interactive elements blend in VTuber livestreaming environments. 

VTubers have the distinguishing factor of being avatarized entertainers. Yet, unlike many prior examples in the literature, VTubers also incorporate a distinctly human factor as well, via the performer --- the underlying voice actor/actress who plays the role. We study how the unique aspect of VTubers, combining the identity of a real, underlying human with a fictional, front-facing avatar, affects traditional interactions and motivations for viewership.  

\subsection{Livestreaming --- Viewer Interactions and Motivations}
Livestreaming, the practice of broadcasting live video in real-time, has become a popular and impactful media phenomenon \cite{thirdwave, lu2018watch}, with a wealth of academic research in the field. One major area of interest involves understanding the motivations of viewers in watching and engaging with streamers. Sjöblom and Hamari considered this question from a uses and gratifications perspective, finding that tension release (the desire for escape and diversion), social-integrative motivations (the desire to form online connections), and affective motivations (the desire for enjoyment) are major motivators for watching video game streams \cite{sjoblom2017people}. Hilvert-Bruce et al. made similar findings, quantitatively demonstrating that aspects of entertainment, social interaction, and information-seeking serve as motivators that affect livestream viewership \cite{hilvert2018social}. Lu et al. found that abstract concepts such as the personality of the streamer and the general atmosphere of the stream are defining factors that affect viewer engagement \cite{lu2018watch}. Wohn found that factors such as the streamer’s appearance, perceived level of interactivity, and parasocial feelings towards the streamer affect the level of social support attributed to a streamer \cite{wohn2018explaining}. Livestreaming has been compared to performance and theatre \cite{li2019live, scully2017playing, liud2021virtual, beyesPerformingDigitalPerformance2017}. We draw from performance studies in our work --- Dixon describes how digitalization and new media are incorporated into performative art; concepts such as liveness, interactivity, and the concept of the ``virtual body'' pervade our findings on VTubers, especially in regard to livestreaming and avatarization \cite{dixonDigitalPerformanceHistory}. Concepts discussed in Dixon's work, such as the sense of communality between live audience and performers, the blend of real performance with digital imagery, and the malleability of virtual bodies, are all relevant when considering this area.  

In particular, interactivity is key to livestreaming, and viewers have many options in this regard, e.g. sending gifts, commenting in chat, joining fan groups, or obtaining memberships or subscriptions \cite{lu2018watch}. Each of these interactions has distinct motivations for the viewer, which have been widely studied through past literature \cite{li2020systematic, ZHOU2019100815, li2021examining, wohn2018explaining}. For instance, Wang and Li showed that aspects of the streamer’s gender, number of viewers, and stream activity could affect audience comments during the stream \cite{wang2020motivates}. The characteristics of appearance (and gender) within livestreaming have drawn significant scrutiny \cite{zhang2019gender, ruberg2019gender}; we expand upon these factors given that avatarization blurs the line regarding who the streamer actually is and what the streamer looks like. 

Li et al. presented a comprehensive review of user behaviour in video game live streaming, summarizing the characteristics that affect and define audience demand and interaction behaviour, among other topics \cite{li2020systematic}. Prior livestreaming studies have not specifically differentiated between VTubers and regular streamers; in this study, we focus on understanding how the defining characteristic of VTubers of avatarization affects these studied interactions.

\section{Methods}

As our research questions placed a heavy emphasis on the experience of VTuber viewers, \textbf{semi-structured interviews} formed the foundation of our methodology. Through these exploratory interviews, we gleaned information about the unique nature of VTuber streams, the motivations for viewership, and the relationships viewers had with VTubers they watched. Findings from these interviews helped build on top of background knowledge and prior literature. We interpret our findings to draw out the specific influence of avatarization on these livestreaming factors. 

To preface, despite the use of the term ``viewers'' in this research, our study also considers activities beyond simply watching the stream --- people may also browse community forums or discuss the streams in smaller local communities. Thus, we define a viewer in this context as someone who may encompass a wide variety of fan-related interactions revolving around the primary activity of watching the stream. 

\subsubsection{Participant Recruitment}

We targeted regular viewers of VTubers (`regular' being up to the interpretation of the participant). To reach potential participants, a call for participation was created and posted on online study listing boards and within targeted Discord communities. To offer a more detailed and focused interview, we contacted participants before the study to ask them for a few videos, streams, or clips of VTubers that they enjoyed watching. Preparatory work was typically performed in terms of reading about specific characteristics of these VTubers (often well-documented and sourced on fan wikis), as well as looking through the sent-in videos to note down the content and themes, especially during popular moments. Two interviews were discarded post-hoc because during the interview, they self-reported as not regular watchers of VTubers, leading to a final analysis of 21 participant interviews. All of the final participants (average age: 20.7, gender: 14 male, 6 female, 1 non-binary) were noted to be active VTuber watchers --- participants indicating that they watched at least around 1 hour of VTuber content per week (with an average of 9 hours a week). 

\subsubsection{Interview Protocol}

Semi-structured interviews were conducted through remote video calls, and audio recordings were made with participant permission (all participants signed a consent form outlining ethics approval before the study). The researchers initially developed questions revolving around the research questions, mediated around prior research in livestreaming and VTubing. These included topics regarding motivations they have for watching, what sort of relationships they develop with the VTuber, how they perceive the performer and the avatar differently, etc. If the participants watched traditional real-person livestreamers, they were prompted to consider the differences in motivation and perception as well, especially factors relating to the difference in avatar representation. As the participants were all viewers of VTubers, the bulk of the discussion centred around the livestream experience; however, conversation points regarding interactions outside of livestreaming (e.g. in online fan communities) were brought up when applicable. Interviews generally lasted approximately 45 minutes, and participants were compensated \$10 CAD for their time. 

\subsection{Data Analysis}

The transcribed interviews were analyzed through a reflexive thematic analysis approach \cite{braun2012thematic} by the primary researcher. Initial open coding was performed to capture the relevant aspects of the data, creating both latent and semantic codes. This initial coding process was fluid and evolving, codes were refined and iterated as we combed through the data. The researcher approached the data both deductively, based on our prior background knowledge about livestreaming, avatar-based interaction, and VTubers, as well as inductively, using user responses to glean deeper insights into novel findings. The primary researcher also aimed for a level of reflexivity within the analysis while being aware of their positionality, offering an informed interpretation of the interview responses within prior background research and their own experiences. Hierarchical mapping techniques grouped these initial codes into broader categories that would help in developing the themes explored in the subsequent findings.

\subsection{Researcher Engagement}

During the time of this study, the primary author made a concerted effort to engage within and observe the VTuber fan community, actively watching VTuber streams with an eye for considering the relationship between viewers and VTubers and noting down moments that differentiated from traditional live streaming. The author also browsed community forums (Reddit, being the most popular discussion board brought up during the interviews) in a structured way --- focusing on three main subreddits --- /r/Hololive \cite{r/hololive}, /r/Nijisanji \cite{r/nijisanji}, and /r/VirtualYouTubers \cite{r/virtualYouTubers} (the first two being official subreddits of VTuber corporations, the latter being an unofficial subreddit for general VTuber discussion). The primary author read posts on these subreddits regularly for a month-long period, noting down posts that reached the front page (i.e. reaching the top 25 in popularity). For both observations of streams and analysis of subreddit posts, the researcher logged critical information in a structured way using coding sheets. Such observations allowed the researcher to corroborate the findings mentioned from the participants' perspectives.

\section{Findings}
\label{sec:findings}

\subsection{Characteristics of VTuber Interactions}

Avatar-based livestreaming provides an altered set of interactions with the audience due to its distinct set of affordances. We explore how stream and community content is affected by the avatarization of VTubers, addressing \textbf{RQ1}.

\subsubsection{Interplay between Realism and Fantasy}

The virtuality of VTubers, drawn from their virtual avatars, allowed them to create and extend upon methods of engagement and interactive effects that could be more ``fantastical'', and thus, perceived to be more entertaining. VTubers could engage in activities that change their model in different ways, for example, by engaging in genderswaps (i.e. changing the model of their VTuber to a different gender, often accompanied with a voice modifier), avatar swaps (i.e. a different performer provides the voice, motion, personality for the usual one), or creating interactable virtual props that would defy real-world physics. This was prevalent in both observed streams by the researcher and also corroborated by the interviewees, e.g. \emph{``VTubers can mess around with their models on stream''} (P11), and \emph{``they’re virtual, they can make anything, like floaty stuff or props''} (P15). These fantastical characteristics added a layer of streamer engagement, manifesting in the creation of fanart, memes, and lore discussion \cite{lore1, lore2}. The virtuality and ephemerality of appearance afforded these fantastical interactions unique to avatarized entertainers --- \textbf{VTubers can take advantage of their virtuality to create effects that defy real-life logic}. 

However, on the contrary, VTubers had difficulties performing in ways that were too grounded in reality. For example, for cooking or crafts streams, VTubers would typically have a camera recorder directed toward the performer’s real hands, while the avatar still appears on the screen. To preserve anonymity and maintain their consistency in having an avatar-forward appearance, popular VTubers might wear gloves and ensure that the camera does not capture their real body. Participants indicated that this was probably difficult --- \emph{``It’s harder because you have to hide everything about yourself, like your hands ... you’re probably gonna have gloves on and stuff like that when cooking''} (P11). In addition, participants mentioned that VTubers also have limitations if they wanted to deliver content that involves traversing the real world (IRL streams) --- \emph{``in Japan, we have a lot of TV shows about people who go in and around the city and buy food and report and stuff ... But that’s not possible --- VTubers cannot do that.''} (P8). Similarly, fan meetups were also discussed as a limitation for VTubers, \emph{``the fan meetups I’ve seen them do is like a TV screen, and the fan on the other side, and that feels a little bit odd to me''} (P20). Thus, a key finding is that for VTubers, who are presented as avatars, \textbf{livestream interactions that require a body representation in real-world space or interactions with real-life objects become significantly more difficult} --- cooking, walking around, and so forth. 

\subsubsection{Escapism --- Fantasy, Absurdity, and Real-life}

Extending on the prior finding, we identify one driving motivation of VTuber viewership as \textbf{escapism}. Participants stated that watching VTubers provided them with a sense of escape from present reality into a different realm, often making comparisons to other forms of entertainment media, e.g., \emph{``It’s sort of like escapism --- it’s just them having fun playing games ... I don’t have to worry about work or other issues''} (P6), and \emph{``VTubers are quite a nice escape from reality. I think a lot of anime/manga ... feels like it can be an escape''} (P12). 

This latter quote discusses escapism through a direct comparison between VTubers and fictional media --- tying towards the otherworldly, fantastical, and whimsical elements drawn from and afforded by the virtual character. Sjöblom and Hamari did present escapism (as part of ``tension release'') as one of the key factors in why people watch gaming livestreams \cite{sjoblom2017people}. However, in comparison to traditional streamers, we found that the diminished level of reality in VTubers (drawn from their fictional avatar, its accompanying design, and its associated lore) could further augment these effects of escapism, drawing a more distinct boundary between motivators as a result of avatarization. Participants often discussed how it was interesting and somewhat unexpected to see anime-styled characters perform in real life --- \emph{``Good analogy my friend uses --- you get to see anime girls do IRL things''} (P4), \emph{``It’s kind of funny... hearing some anime girl saying some stuff that is low-brow, not really sophisticated''} (P5). The \textbf{absurdity} and comic nature of watching (usually) anime characters come to life distills down the unique entertainment aspects of VTubers, forming the main idiosyncrasy that differentiates them from traditional livestreamers. This absurdity might further diminish the feelings of realism during the stream, blur the line between fiction and authenticity, and contribute to escapism arising from the incongruous blending of a seemingly fictional character into the real world. 

However, escapism was also facilitated in a more grounded way as well, tying into the entertaining interactions that exist between streamers. A majority of participants indicated a big draw of watching VTubers was their collaborative streams involving various interactions between different VTubers. In these collaborations, multiple VTubers occupy the same stream space (e.g., by allowing their avatars to use the shared stream video feed or using a mediating technology such as social VR). Thus, despite their performers potentially being physically distant, collaborative streams give the perception of visible closeness, which is only possible through avatar representations in virtual space. Perhaps as a result of this visible shrinkage of physical distance, participants highlight the group intimacy of collaborative streams where \emph{``I think the escapism appeal of [VTubers] is like having what seems to be an incredible group of friends''} (P12) and \emph{``I think it’s just that watching the dynamic between the streamers is pretty fun''} (P1). This facet of group identity was discussed by some participants as being a benefit of VTuber corporations. This tied into the feeling of being in almost a familial group of friends, a sentiment echoed both by VTubers and viewers alike, e.g.  \emph{``There’s a lot of different kinds of interactions that they can play off each other well during collab streams, I just enjoy that a lot''} (P11). As such, an additional facet of escapism also arises from a quite grounded, real-life desire for \textbf{social collectivism and community} --- extending individual entertainment into a feeling of communal camaraderie. 

\subsection{Audience-VTuber Relationships and Motivations for Viewership}

Although the prior findings apply to VTubers as a whole, we asked interviewees to further describe their favourite VTubers and what sets them apart from others, providing the foundational basis to address \textbf{RQ2}. The word ``oshi'' was used as a term that refers to the dedicated viewers and fans of a specific VTuber. Numerous different characteristics were discussed, and in differentiating VTubers from each other, we separated these motivators into the two categories of \textbf{avatar-related factors} (dimensions that relate to the avatar of the VTuber) and \textbf{performer-related factors} (dimensions that relate to the human playing the VTuber). 

\subsubsection{Performer-Related Factors}

Several performer-related factors strongly affected a viewer’s desire to watch the VTuber. Most importantly, every interviewee described the personality of the VTuber as a main factor that drew their attention and continued viewership --- \emph{``[I] also like [Gura’s] personality because she’s, I guess, kind of sassy''} (P3), \emph{``there’s so much I love about [Kanae’s] personality''} (P12), \emph{``I’m not sure if I can put this into English, but it seems like [Ina’nis] has this calm vibe''} (P17), etc. --- personality often viewed more of a representation of the underlying human. Thus, the most important factor in which VTubers resonated most closely with their audience was compatibility in personality (\emph{``If I want to watch [any VTuber] again, it’ll be for their personality''} (P2)), which was a largely subjective factor that differed across all viewers. In addition, other performer-related factors, such as their interests, voice, and languages, were additional factors that affected how much a user liked a specific VTuber. This was particularly apparent when it was a shared characteristic with the viewer, e.g. \emph{``I really admire [Suisei’s] singing skills because I’m personally into singing a lot and she’s really a good singer''} (P8), or \emph{``[Ironmouse] speaks Spanish, which I also speak''} (P5). 

\subsubsection{Avatar-Related Factors}

Avatar-related factors also affected a viewer’s enjoyment of a specific VTuber. Interviewees brought up how design and appearance could be selling points for many VTubers, e.g. \emph{``[One of] the [VTubers] I watch the most would be Subaru from Hololive ... I just think she’s cute''} (P1), \emph{``character designs are also part of the reason why I like certain people like Gura''} (P3).

Gender was a cue that draws initially from the avatar’s appearance (as well as the VTuber’s voice). The apparent gender of the VTuber played a role in a user’s motivation for watching --- the vast majority of popular VTubers are female. From a historical perspective, interviewees mentioned that this was generally a stereotypical reflection of the initial targeted demographic who would want to watch avatarized anime livestreamers \cite{reysen2016pale}, also mimicking prior studies that studied VTuber viewers \cite{lu2021kawaii}. As such, although the majority of participants indicated that they largely watch female VTubers, several stated that this was just a result of initial exposure since female VTubers form a vast majority and have been historically more prominent. However, the release of new generations of corporate male VTubers has caught the attention of many viewers --- illustrating an emerging audience for male VTubers as well, a sentiment echoed by the participants --- \emph{``it’s pretty clear that there is an audience for male VTubers as well''} (P11).

Tying into gender, a few participants mentioned that gender acts as a motivator for them in watching certain VTubers, e.g. \emph{``I definitely feel like there’s this kind of interaction of like, I like this person in not the way I would like a man''} (P8). Viewers indicated that \emph{``I think basically all the guy VTubers I like, they’re all cute ... I like the cute aspect of it''} (P12), and \emph{``I think part of it is like a lot of guys want to see cute anime girls''} (P11). We interpret this as specific gender-based motivations stemming from feelings of attraction. The primary researcher's observed engagement with livestreams offered further proof, in which chatters would comment on the appearance of the VTuber, often in a praising manner (e.g. ``pretty'' or ``cute''). The term ``cute'' was brought up very frequently in interviews, as well as the researcher's observation of Reddit threads and of livestream chat. The factor of ``moe'' borrowed from otaku subculture, describes a concept relating to the desire for certain aspects of cuteness \cite{fandomunbound, azuma2009otaku}, however, this concept has garnered increased widespread use outside of its natal subculture \cite{marcus2017cuteness, zhang2020cognitive, yiu2013kawaii}. We highlight how this plays a similar role in the appeal for VTubers.

\subsection{Perceptions of Avatar Appearance and Design}
\label{sec:appearance}

Avatarization creates the idiosyncratic draw of VTubers and most distinctly affects their appearance, the aspect that most blatantly sets them apart from other livestreamers and media personas. Here, we specifically discuss the impact of avatarized appearance and design on viewer perception and interaction. 

\subsubsection{Appearance as an Equalizer}

We first find that \textbf{avatarization provided a method of equalizing appearances}. The virtual nature of avatarization allows VTubers to design themselves to look ``good'', ultimately allowing everyone to look as appealing as they desire. An attractive appearance is a large draw for avatar loyalty \cite{li2018avatar}, and participants stated that for VTubers, \emph{``You always look nice because you’re designed that way''} (P11). Avatarization also prevents viewers from judging VTubers based on real-life standards of appearance, e.g. \emph{``I can’t judge them if they have a strange room, or they look weird ... Like I can’t ignore that for someone else''} (P9). As such, because appearances felt more equal to the participants, this placed an increased load on the personality of the VTuber as a differentiating factor, \emph{``I think the thing about VTubers is that the personality of the [performer] is more important ... it’s less focused on appearance''} (P5).

\subsubsection{Appearance as a Differentiator}

Although the visuals were more equal in terms of the objective ability to look appealing, \textbf{the visual designs of avatars were also differentiated by their subjective appeal}. To preface, all of the VTubers that participants mentioned watching were largely anime-styled. Yet this is not a necessary characteristic of VTubers, as any humanoid model can be motion-captured. However, this anime style was a general appeal for viewers, as most participants indicated their past or present familiarity or engagement with otaku culture --- \emph{``that’s definitely a huge draw for me. I think the Japanese anime, manga art style is really appealing''} (P12), and \emph{``obviously, a huge factor of why VTubers are appealing is people who watch anime probably are going to like VTubers because of a similar aesthetic''} (P11). Thus, the anime art style prevalent among current VTubers served as one factor within the concept of subjective appeal. A differentiation was drawn among specific VTubers as well, in which participants indicated that they like certain VTubers because of their models, e.g. \emph{``I just like the design of Gawr Gura the character''} (P3). Viewers indicated that the same performer with a different model may not be as appealing to them, e.g. \emph{``the same personalities with a different rig ... I might not get into them''} (P12), and \emph{``even if that person has a really great personality ... I feel like that is something that wouldn’t be able to appeal to me''} (P21). Thus, we highlight that even though appearance is something that VTubers can choose for themselves, it still acts as a subjective factor that affects individual viewer motivation. 

\subsubsection{Appearance and Avatar Sexualization}

As established, appearance can be a motivator for viewership, with chatters often commenting with flattering adjectives to describe VTuber appearance. However, sexualization (based on appearance) was also apparent within the VTuber community and the interactions within it. Although such factors exist and form areas of controversy within real-person streamers as well \cite{zhang2019gender, ruberg2019gender}; \textbf{manifestations of sexualization differs with an avatar}. One way in which sexualization is apparent is through fan creative engagement with the VTuber avatar, e.g. NSFW (not safe for work) fanart was the main way this occurred. In these cases, the sexualization applies solely to the avatar and is largely disparate from the performers themselves. This was largely accepted by the community --- comments that the researcher noticed under NSFW fanart might sometimes even be largely positive towards the art. 

Fanart appeals more towards the avatarization side --- it captures the appearance of the avatar of the VTuber, rather than anything related to the performer. Due to the distance between the performer's identity and the avatar's identity, it perhaps seemed more disparate from affecting the actual human. On the other hand, sexualization through livestream comments, which may directly interact with the performer, was seen to be less acceptable by the participants. Although the VTuber’s reaction may differ, a few participants perceived this negatively, calling it \emph{``cringe or like, kind of gross''} (P5), showing that there is a perception that some viewers act inappropriately regarding sexual motivations. 

\subsection{Relationships with the Avatar, Character, and the Performer}

We find that aspects drawing from both the performer and the avatar affect the motivations for watching VTubers, diving deeply to address \textbf{RQ2}. We consider how these two separate identities --- one virtual, one real --- work together to create the singular entity called the VTuber, highlighting how the mixing of identities affects the audience-facing experience of viewership. All interviewees were aware that these two identities exist separately. Knowing this, we consider how audience members perceive and develop relationships differently regarding each of these separate identities. 

\subsubsection{The Role of Role-Playing}

VTubers use avatars as their outward-facing appearance, often accompanied by fictional lore and history. Participants presented a variety of reasons for these unique, often-fantastical avatars that are disparate from reality, such as wider freedom of imaginative engagement, e.g. \emph{``It gives them more room to express or more room for imagination in a way ... more ways to interact with the audience''} (P6), \emph{``they can have more fantastical designs ... you can have lore that comes with it --- that could be just an immediate thing that pulls you in''} (P11). This is represented in the methods of engagement outside of streaming as well --- the avatarization and fantastical nature of avatars tie into numerous pieces of fan art, discussions on lore, and memes often found on the associated subreddits, opening up another avenue for fan interaction and reflection. Thus, the fantastical elements of VTuber characters added a layer of interaction and engagement for the viewers, increasing audience investment as they can interact with the avatarized ``character" as a disparate entity from the identity of the streamer. Thus, for VTubers, \textbf{the character associated with the avatar continues to exist in perpetuity beyond the livestream}, offering a unique sense of continuity. 

\subsubsection{Blending the Real and the Virtual}

Participants discussed how VTubers blend the identities of the performer and the fictional avatar to build a coherent, consistent whole. The perception of this blending was different across all VTubers --- whereas some VTubers tie very much into acting the character, others lean towards representing the performer, often discussing aspects of the performer’s personal life and story through the medium of an avatar --- \emph{``there’s a very clear spectrum of like, character playing to [playing] based on your own personality''} (P14). Overall, it was up to the VTuber's discretion on how performative they wanted to be with their presented character. 

To some extent, this blending required a level of perceived agreement between the avatar and the performer. When the blending felt natural, viewers viewed the VTuber more coherently as a singular existing entity; on the other hand, viewers described certain other VTubers as a performer trying too hard to fit themselves into the mold of the avatar, which could sometimes seem overbearing in their opinion, e.g. \emph{``Gura and Ina were pretty natural about it. It just kind of merged their own personality with their character. And then they tried to just mend and mesh the parts that don’t fit too well together''} (P2). Ultimately, viewers indicated this blending was up to the performer’s personal decision --- \emph{``it’s up to the creator to find a balance between how much of their real-life they want to let in or keep out''} (P3). However, the fact that the blending does not always seem to mix well suggests that there is a conflict between what people expect the performer to be like (based on the avatar) and how they are, implying that \textbf{the avatar itself predisposes some idea of the character}, even with no human performer. 

The duality of VTuber identities creates unique and memorable moments within livestreams in which the performer fit their personal stories into the lore of the avatar, e.g. \emph{``Pekora talking about her mom, right? ... she’ll also talk about how her mom is like part rabbit or something ... keep it a little bit in character also''} (P5). However, the general trend that many participants mentioned was that at the start of a VTuber’s career, they typically play more of the avatar character --- \emph{``the general trend I noticed is it falls off over time. So the first day they debut, they follow really strictly to the character''} (P4), and \emph{``after like a couple months, ages, it blends in more of their actual personality more and more''} (P1). Participants agreed that they would find it difficult to act a fully distinct character for several hours daily --- \emph{``if you’re streaming, it’s really hard to put on a completely fake persona. So the best you can do is to kind of combine the character''} (P2). This highlights that one of the unique challenges for VTubers is understanding to what extent they would like to play a character versus to what extent they would like to be authentically themselves. This raises an important question, however --- if a streamer wants to be fully authentic, what purpose does an avatar serve? Therefore, we argue that \textbf{using an avatar almost inherently involves some level of blending of identities and character portrayal}. 

\subsubsection{Engaging with the Performer}

Motivations for watching specific VTubers involve both aspects of the performer and the avatar character. Similar to how aspects of the performer's personality become more definitive over time, most participants indicated that performer-related motivational factors also become more prominent and important to them over time. Appearance and design were noted as important ``hooks'' to get people interested in watching, but over time, the personality and charm of the performer demonstrated through the medium of the avatar is what kept them watching, e.g. \emph{``you’re here for the character until you are here for the person and not the character''} (P11).

Although the unique engagement and interactions born from fantasy offer moments that are fun and entertaining, participants found that the rarer times in which performers discussed aspects of their own life felt comparatively more personal, relatable, and important, e.g. \emph{``like you can sort of relate to the character''} (P6). Identification and authentic intimacy have been described as key motivators for livestream watching \cite{hamilton2014streaming, lu2018watch} and challenges for virtual influencers \cite{louAuthenticallyFakeHow2023}, and their scarcity among VTubers' streams exacerbated their emotional effects. Participants expressed sentiments such as \emph{``here’s a person talking about more personal things and that can be not just entertaining, it can make you feel nice emotions... It makes it feel a lot more personal and a lot more important''} (P11). While aspects of avatarization may be important initially, they may become less crucial over time as bonds and relationships solidify and the inherent nature of the VTuber shines through.

This also demonstrates that \textbf{both the realism afforded by the performer and the virtuality afforded by the avatar have distinct roles in establishing interactions and relationships}. Methods of fan engagement tie into both; for example, we found that on the subreddit, a VTuber will sometimes be discussed in regard to their realistic qualities (e.g. by their country or the real-life activities of their performer) and other times in terms of their fantastical qualities (e.g. by their lore). 

\subsubsection{An Increased Sense of Distance}

Participants indicated that the public-facing avatar combined with the common anonymity of VTubers \textbf{creates a barrier in developing parasocial relationships with the human identity of VTubers} (i.e. the performer) when compared to real-person streamers. Participants mentioned that \emph{``I think it’d be even worse to think that you have a parasocial relationship... there’s an additional wall between you and a VTuber because you don’t even know what they look like''} (P2) and \emph{``You don’t know a person quite as well as you would with an IRL streamer''} (P1), indicating that viewers felt an increased sense of distance. This was mirrored through other engagement mediums such as Twitter --- \emph{``[Pekora]’s gonna tweet about not anything related to her actual life ... So [fans] feel less connected in a social aspect''} (P5), and corroborates prior studies regarding VTuber-viewer interactions \cite{lu2021kawaii}. 

Numerous factors contributed to this increased feeling of distance. The lack of a human appearance was one major factor --- when describing the differences, participants indicated that for a real-person streamer \emph{``You can see their face to talk about their life and everything''} (P1) and \emph{``it’s very much you’re closer to the person as an individual, because you see a lot of their outside content, and you see a lot of themselves in their content''} (P14). The usual anonymity of VTubers was another factor, e.g. \emph{``if you keep a level of anonymity ... when you’re trying to relate to the viewers or when the viewers are trying to relate to you... it doesn’t hit all the way''} (P21) as well as aspects of fantasy and lore \emph{``avatars aren’t like your average Joe, right? You’re not gonna always relate to this character''} (P21) also played a part in the increased sense of distance that viewers felt when watching VTubers. All of these facets can relate to the avatarized nature of VTubers, highlighting how avatarization can foster an increased sense of distance in streamer-viewer relationships. Participants did not explicitly denote any difference even in the rare occurrences in which they were aware of the VTuber's identity or previous online pseudonym, which could come up e.g., \emph{``by accident, I was listening to a song... and I was like wow she sounds really familiar''} (P12) or just through \emph{``random [Youtube] recommendations''} (P2). 

However, perhaps surprisingly, a small number of the participating viewers stated that this increased sense of distance was preferable to them. One viewer indicated that \emph{``It is sort of comforting for me because I don’t have to fully engage in supporting that person... It is an avatar, so I can be passively engaged towards the video, I can reap the benefit of watching without committing anything''} (P8), indicating a sense of consuming the content without needing to feel committed or related to the streamer. Similarly, P12 indicated that \emph{``I suddenly think about what they look like, or what they are like as humans, and that’s kind of weird for me. I like having the disconnect''} (P12), showing a desire to maintain separation between reality and entertainment. We find that for certain viewers, the engagement and commitment associated with parasocial relationships often found in livestreaming are things they actively aim to avoid and prefer not to have. As such, these viewers can find solace in watching VTubers, whom they perceive to have a lessened parasocial effect.

\section{Discussion}

We found that aspects of both the real performer and the virtual avatar affect the motivations of, relationships with, and perceptions by VTuber viewers. As performative entertainers that combine both the virtual and real, VTubers fall into a space that is neither fully one nor the other. Building on prior research on technological philosophy, our study on VTubers harks to Ihde's three bodies --- the physical, real-world body, the socially and culturally constructed body, and a third body that exists interactive with technology \cite{dixonDigitalPerformanceHistory}. This unique position they occupy offers them a distinct set of affordances and limitations that differ from traditional livestreamers. 

We first discuss how the nature of avatar-based livestreaming intersects existing HCI literature in system-based livestreaming implementations. This could potentially open the door for systems that mix novel inputs with programmable outputs (the avatar) to create new interactions that have never been seen before. Contextualized around experiential research, we discuss how avatar-based livestreaming affects the prior-studied motivational and relationship factors in terms of livestreaming viewership. We highlight the evolving dynamics in these factors that an avatar brings when it comes to livestreaming, which can inform decisions regarding avatar-based human agents for future HCI designers and researchers. We finally consider how avatarization might affect viewer perceptions of identity and how users might perceive, interact, and form relationships with avatars with blended identities. 

\subsection{Avatarization Extends Livestreaming Engagement in Novel Ways}

\textbf{Avatarization affects the space of interactions between VTubers and viewers}, especially in comparison to non-VTuber streamers, and we highlight how this may 1) affect motivations and 2) extend into further novel interactions. The key characteristic of VTubers that causes engagement to shift is their virtuality in appearance --- VTubers appear as avatars rather than real humans. Within our findings, we found that this aspect of virtuality first allows for degrees of fantasy and whimsy; by engaging with the virtual and taking advantage of the ephemerality of their public identity, VTubers were able to develop forms of entertainment that go beyond traditional livestreaming. This creates novel forms of escapism and fantasy that non-VTuber livestreamers cannot achieve. Tying this into past studies on motivations for livestreaming viewership, this offers an increased sense of novelty \cite{lu2018watch} and entertainment \cite{hilvert2018social} for viewers, both leading to increased engagement for viewers. From Sjöholm and Hamari’s work of integrating livestreaming into use and gratification perspective, these unique activities of VTubers tie into aspects of tension release (relating to escape and diversion) and affective (relating to enjoyment) needs \cite{sjoblom2017people}. 

Researchers are always looking to extend livestreaming interactions in fun and novel ways to increase engagement and motivation for viewership as well as get the audience more involved, for example, providing users with a visual polling system \cite{chung2021}, using heart rate as input \cite{robinson2022}, and so forth \cite{lu2021streamsketch,  Hammad2023MARS}. Whereas these novel inputs have typically been studied in affecting the software side of systems, we can extend these findings to directly impact the virtual avatar itself (and the associated props), as the virtual avatar is a programmable element that can take input and be altered. As an example, interactive avatar modifications, such as influencing VTuber appearance in real-time through audience input (such as accessories, hair colours, etc.) could be accomplished through existing options such as chat commands or donations. To extend on this, what if platforms also integrated features that could use physiological or behavioural inputs (e.g. heart rate, as in Robinson et al.'s work \cite{robinson2022}) to fundamentally affect the streamer avatar? What novel interactions might this afford, and how might this level of control that the audience has over the avatarized appearance affect the audience's relationship with the streamer? We have already explored some possible interactions in this paper, such as swapping avatars or changing one's appearance. Future design and research explorations could extend the potential novel ways for VTubers (or avatars in general) to utilize their virtual body representation in fun, interesting, and fantastical ways and better understand how these experiences compare to non-virtual experiences in terms of motivations for and emotions during viewership. This would extend the already malleable virtual body of VTubers, embracing the concept of ``impossible anatomies'' as a form of digital performance \cite{dixonDigitalPerformanceHistory}. We draw inspiration from the pursuit of conceiving engaging content that uses novel digital technologies, where artists and performers have long experimented with technologies that blur the boundaries between real and physical \cite{dixonDigitalPerformanceHistory}.  

\subsection{Avatarization Complicates the Relationships between Viewer and Streamer}

Building upon the idea of body ownership, we explore how the lack of a coherent and real identity for VTubers could correlate with a stronger separation between the avatar and the performer --- as discussed in the findings, this relates to a more distant relationship between streamer and viewer. We consider how an avatar affects the perception of VTuber identity and how it uniquely affects livestreaming relationships. 

Throughout our findings, we found a disconnect between the user perception of the avatarized character and the user perception of the performer, and the relationship that users had with these two represented identities was different. A key tension arises for VTubers: How much does an avatar represent themselves (as an extension of their identity), as in e.g., social VR \cite{freemanMyBodyMy2020a}, and how much does the avatar represent a performative character that they portray? This tension is particularly apparent in how the owner of an avatar talks about their avatar concerning appearance and expressiveness --- whereas VTubers might overinflate their avatar's traits and use the avatar to play into sexual suggestiveness and exaggeration \cite{wanInvestigatingVTubingReconstruction2024}, social VR users might perceive their avatars as having a more intimate relationship with themselves, perceiving judgment on the avatar as judgment towards themselves \cite{schulenbergCreepyMyAvatar2023}. 

Our study extends this tension from the viewers' perception --- a dichotomy was perceived between the authentic identity of the performer and the performative identity of the avatarized character. Perhaps, as people felt more distant from the former, they related to the avatarized character in ways more similar to other forms of media --- making fanart \cite{manifoldFanartCraftCreation2009a}, talking about lore \cite{corneliussenDigitalCulturePlay2008}, forming a relationship with the character \cite{jonesHarbourStrongFeelings2020a}, and so on. We interpret VTuber avatars to fall somewhere between the spectrum of full authenticity (avatars as a medium for self-identity to represent yourself) and full performance (avatars as a medium for performative acting to appeal to an audience). Our study reinforces another constraint --- VTubers have both expectations to play a character as well as expectations to be themselves. Viewers can be initially hooked by a VTuber's attractive design and appearance, and VTubers can freely present their avatars in a `cute' way for broad appeal. Wan et al. find that VTubers may also overinflate their behaviour in a way to further extend this illusion, effectively acting as a filter (but also blurring the line) between their real identity and the virtual persona \cite{wanInvestigatingVTubingReconstruction2024}. Then, \textbf{to what extent does the design and identity of their avatar reflect an authentic extension of their imagined self and personal identity} \cite{freemanMyBodyMy2020a}, and \textbf{to what extent do they represent an optimized tool for mass appeal, popularity, and viewership}? This tension likely arises due to the viewer's perspective. Appearance and visual cues in the form of facial features \cite{paunonenFacialFeaturesPersonality1999} or clothes \cite{sharmaClothingBehaviourPersonality1980, boomsmaFabricatedSelfRole2020} convey personality traits that shape viewer expectations, adding additional factors in choosing between appeal and appearance-personality consistency versus personal desire and self-expression. Inconsistency between avatar appearance and performer may account for instances in which the viewers perceived a `mismatch', which may tie towards a similar idea of misrepresentation (based on an idealized perception) of an avatar \cite{ramadanAIAvatarsCocreation2025}.

For the VTuber, the balancing act of identities \cite{wanInvestigatingVTubingReconstruction2024} may not always be easy, and future research must consider this duality in identities and how people form relationships with each of them. When the VTuber faces a profusion of livestream comments calling them `cute', `pretty', etc. --- who exactly are the viewers really talking about? And what effect may this have on the streamer's perception of themself? How might this possibly tie into the nuanced avatar-self relationship denoted by the Proteus effect for social VR \cite{freemanBodyAvatarMe2021a} and psychological effects for performative acting \cite{burgoyne1999impact} --- how might the actual behaviour or personality of the streamer change when their avatar is constantly complimented (or possibly criticized)? These answers can guide designers in \textbf{finding affordances to aid in fostering healthy connections and clearer boundaries of relationships with avatars that mix a crafted person with a real self}. For instance, built-in stream and platform cues that demarcate between their performer identity and their avatar identity could more clearly communicate to the audience which identity the VTuber is drawing more from. VTubers could also reflect this in their virtual appearance - integrating appearance, features, or stream content that tie into the more fantastical, performative, and lore-relevant when they are drawing more from their avatar identity and tying their stream more towards their authentic self when drawing more from their performer identity. Regardless, we highlight the importance of respecting VTuber autonomy in design, letting them portray what they feel is their desired representation of themselves, irrespective of viewer expectations. Given a dramaturgical perspective of life as performance \cite{smithDramaturgicalLegacyErving2013}, these implications may also extend even beyond VTubers --- as real-life streamers themselves may adopt a carefully curated persona \cite{jacksonUnderstandingMemeticMedia2021}. 

\subsection{Motivations in Viewership are Mediated by the Avatar}

The lack of a coherent identity for VTubers has previously been shown to correlate to lower social presence of the streamer, associated with parasocial interaction and attraction \cite{chinchillaVTuberStreamersExploring2024a}. This has been a common finding regarding virtual influencers at a broader scope \cite{ghzaielWhenArtificialRevolutionizes2025, louAuthenticallyFakeHow2023}. The parasocial relationship present within livestreaming can have strong influences on the viewers themselves, affecting viewing habits and attitudes \cite{wohn2018explaining}, but can also potentially lead to unhealthy effects \cite{jarzyna2021parasocial, webster2019parasocial}. Lim et al. showed that strong parasocial relationships can result in continued engagement and viewing of a livestreaming \cite{lim2020role}. In addition, past studies have shown that authenticity and relatability are two factors that create a strong sense of parasocial relationships \cite{webster2019parasocial}; however, the lack of these facets within VTubers serves as a deterrent to the creation of these relationships --- as VTubers are very blatantly not real people (other facets, like appearance-based motivations, may still exist). As such, the nature of parasociality differs fundamentally for VTubers; our study suggests that viewers feel more distant from VTubers because of the separation brought forth by avatarization; however, these qualitative findings somewhat contrast prior study on parasocial interactions between real and virtual influencers \cite{stein2022parasocial}. Although Tan found that a closer parasocial attachment to VTubers was correlated to stress relief \cite{tan2023more}, in our study, a small number of viewers indicated that they actually preferred this more distant relationship --- this may suggest that VTubers open up livestreaming viewership from potentially new viewers who may want to avoid such emotional engagement within their entertainment (and differentiates the motivations from consumer-based studies on virtual influencers \cite{ghzaielWhenArtificialRevolutionizes2025}). Along the spectrum of media entertainment, if we interpret `livestreaming' to represent an authentic display of a real person and `acting' (e.g. in a show or movie) as embodying a character, then \textbf{VTubers fall somewhere in between and can benefit from motivations from both paradigms}. 

VTubers exhibit a duality of real and virtual identities. Whereas Wan et al.'s work discusses the VTuber's affordances based on this duality \cite{wanInvestigatingVTubingReconstruction2024}), we find that this duality also creates a set of distinct motivations for VTuber viewership. The first of these motivations was performer-related, tying into the personality of the actor or actress of the VTuber. This set of motivators relates more closely to past literature on generalized livestreaming; for example, Lu. et al. highlighted aspects like the streamer's personality and skills as one of the aspects of engagement --- these relate to the streamer’s sense of humour, their talents, and their personality \cite{lu2018watch}. On the other hand, the second set of motivators, the avatarization motivators are a \textbf{distinct characteristic of VTubers}, and form more of a detached persona that has a level of continuity beyond the livestream (i.e. similar more to a movie character), that people interact with beyond the performer; forming the basis of a broader community. Depending on the type of relationship they may want to cultivate, VTubers can lean more into either identity. 

The avatarization motivators tie in closely to past research on streamer appearance. A viewer’s engagement in the stream can be affected by the ambience, streamer personality, and so forth \cite{lu2018watch}, yet, the appearance of the streamer is also a factor that affects viewership, parasocial relationships, and engagement (sometimes with less than savoury intentions) \cite{lu2018watch, mclaughlin2021predictors, ruberg2019gender}. For VTubers, this forms a paradoxical dichotomy because they have full control over their appearance and thus full control over this motivating factor. Yet, we find that, whereas viewers still have subjective preferences for the designs of VTubers, the virtuality of appearance can serve as an equalizer for this motivational factor --- it allows VTubers to design themselves to look however appealing they want --- which can serve as an appeal for viewers but also serves less as a distinguishing feature of any specific VTuber. When all VTubers can choose their design, the general appearance-based motivation of VTubers may be less prevalent when compared to their non-VTuber counterparts. Regardless, gender and attraction were reasons we found that people watched certain VTubers; Wan and Lu highlight how gender stereotypes are often exaggerated to encourage viewership and interaction \cite{wanInvestigatingVTubingReconstruction2024}; perhaps one reason this might be more acceptable to VTubers is the disparity of identity between their virtual and real selves, tying back again towards whether the VTuber represents an extension of self or an optimized tool for marketing. With the malleability of VTuber appearance, VTubers may want to explore what self-presentation works best, balancing between stream popularity, viewer engagement and interaction, and their personal self-expression and desires.

All in all, this section outlines how \textbf{the idiosyncratic characteristic of virtual avatarization of VTubers warps viewer motivation, perception, and the formation of audience-streamer relationships during livestreams in unique ways}. Potential future explorations could focus on how specific aspects of VTuber avatars (e.g. the level of realism, expression mimicking, etc.) might affect such factors and how this might compare to non-VTuber streamers as well. Furthermore, with the rise of generative AI and LLM technology, it might be possible in the future to study the effects of avatarization in isolation. This could be done by using an AI agent as the `human' identity of the VTuber, as there have been examples of VTubers that claim to be created without a `true' performer \cite{neuro}. The implications of this research are increasingly important as avatarization representations become more prevalent in research and industry. Avatars are increasingly being used in fields such as healthcare \cite{shakedAvatarsVirtualAgents2017}, marketing \cite{sakumaYouTubersVsVTubers2023}, and learning \cite{petrakouInteractingAvatarsVirtual2010, falloonUsingAvatarsVirtual2010}. As such, it is important to understand how people perceive, interact, and build relationships with avatars and whether the explicit knowledge of whether a separate human with a separate identity is controlling the avatar affects these factors.  

\section{Present Challenges and Future Explorations}

Finally, we consolidate our findings to highlight the present challenges brought up in the findings and propose future avenues for exploration stemming from these challenges, addressing \textbf{RQ3}.

\subsection{Bridging the Gap between the Virtual and the Real}

VTubers can have trouble engaging in ways that are too grounded within reality --- they have difficulty partaking in activities that involve objects in the real world while maintaining the consistency of avatarization. Future explorations could investigate how a virtual character could interact with real-world objects in real-time. One such potential area could involve investigations into mixed-reality interfaces that inherently blend the real and the virtual. For example, augmented reality has been used to involve VTubers in real-world locations but lacks the aspect of the real-time core to interactive livestreams \cite{suisei1}. As the core aspect of livestreaming is the viewer-facing video, designs should also consider the presentation of such activities onto a video feed and how the presentation maps toward real life. 

Viewers, also being inhabitants of the real world, are limited in the ways they can interact with VTubers. Currently, virtual reality has been applied as a solution to the content limitations of VTubers because it allows the viewers, acting as the ``real'' users of the technology, to interact within a shared virtual world (e.g. in social VR-type settings \cite{divinefalling}) --- bridging the divide between VTubers and viewers. Such spaces act as a common space facilitating direct interaction that can draw from a diverse set of interaction cues \cite{maloneyTalkingVoiceUnderstanding2020b, fangSocialInteractionsVirtual2021}. In such situations, viewers can interact with streamers with an increased sense of direct presence and embodiment \cite{freemanBodyAvatarMe2021a}, and it may be interesting to study how the perception of ``distance'' changes in such contexts. This may allow for more intimate and novel forms of interaction; we can perceive this as being similar to meetups in the real world, where people can interact with their favourite VTubers outside of being an audience member of a livestream. Yet, challenges such as unwanted harassment due to the physicalized nature of social VR and the possible invasion of personal space \cite{freemanDisturbingPeaceExperiencing2022} exist. Areas of exploration may include designing security measures or enforcing personal space in social VR, akin to Bönsch et al.'s work \cite{bonschSocialVRHow2018}.

Beyond just VR, future research could focus on additional solutions to address the discussed limitations of VTubing content through \textbf{creating novel methods of merging virtuality and realism}, allowing more forms of direct interaction that are currently inhibited by the divide between virtual avatars and the real world.

\subsection{Avatar Ownership and Virtual Replicability}
The avatars that VTubers use are virtual assets, and thus that induces several unique issues. For example, one topical issue is that of belonging --- who does the asset belong to? Several stakeholders are involved in the creation of a VTuber model --- the illustrator who creates the art for the model, the modeller who rigs the model for motion capture, the performer who breathes life into the model, and optionally, the corporation who oversees everything. These stakeholders often work together and maintain friendly relations (for instance, VTubers often call their illustrators some variant of ``father''/``mother''). However, issues concerning ownership can result in problematic scenarios like copyright claims against the use of an avatar \cite{melodycopyright}. Another problematic scenario may occur if a performer is released from a corporation and is forced to relinquish their avatar, which to them represents their entire public-facing career as a livestreamer. As avatarization, in general, becomes a more pronounced concept in real environments, one important area of research may revolve around \textbf{copyright policy regarding what constitutes ``owning'' a virtual avatar} \cite{ochoa2011owns, marcus2007fostering}. 

Due to the virtual nature of avatars, they can be replicated. Thus, it may be possible for bad actors to take advantage of the public trust that the community has in VTubers by using a similar model to impersonate them (e.g. for illegal activity), a hypothetical scenario brought up by one of the interviewees. As machine learning and artificial intelligence continue to evolve, it may even be possible to use a bot to control these replicated avatars. Thus, another area of research may involve exploring \textbf{methods to distinguish human operators of an avatar from bots or other human avatars}, i.e. extending anti-spoofing research as seen in speech imitation \cite{wu2016anti} or facial recognition \cite{akhtar2017face} to avatar dimensions. 

\section{Limitations}

Our sample interviewees skewed towards younger males. Although this demographic skew was also present in a prior VTuber audience interview study \cite{lu2021kawaii}, we emphasize the need for future work with a more balanced gender spread, especially given the studied effects that gender roles have on motivations and relationships during livestreaming \cite{todd2017gender, choiViewershipHowStreamerViewer2022, kneiselEffectsParasocialAffinity2022}. Furthermore, our work also reflects the perspective of primarily a Western demographic, which contrasts prior work which focused on VTuber viewership in China; this is important given differences in cultural perspectives on livestreaming \cite{guanWhatInfluencesPurchase2022, wangResearchMotivationViewer2024}. Increased diversity will help with both enhancing the generalizability of findings and also provide deeper insights into the variation of perspectives across audience segments. With a more balanced demographic and larger sample size, we propose quantitative or mixed-methods studies that investigate motivations and engagement (e.g. using metrics such as the emotional connectedness scale or time spent watching \cite{hilvert2018social}) to strengthen and compare against our exploratory qualitative findings.  

In the interviews, every participant brought a different experience, resulting in a vast number of different VTubers discussed in our interviews as well as different patterns of consuming the entertainment. As VTubers themselves all act differently, it becomes hard to generalize the present set of findings to all viewers and VTubers. Additionally, while every interview participant was a viewer, not everyone was an active participant in the online community. Thus, our findings lean heavily towards the audience members' perspective towards VTubers as primarily livestreamers, and thus, community dynamics regarding VTuber space could be explored in future work, focusing on the dynamic interactions within the broader fandom. 

Furthermore, there was a minimal number of independent VTubers discussed during interviews --- the vast majority belonged to a corporation. Thus, some of these findings may be more esoteric towards corporate VTubers. One possible way to standardize the experience among viewers and to incorporate independent VTubers would be to provide a shared list of VTubers for participants to watch before the study, but limitations and respect for the interviewee’s time prevented us from pursuing this approach. However, in the future, we could recruit a more diverse sample of VTuber demographics and platforms. Furthermore, a study that allows for interviews with the performers themselves (both corporate and independent), such as in Wan and Lu's work \cite{wanInvestigatingVTubingReconstruction2024} or staff from management companies, may help provide deeper insights and validation of the findings discussed. This would help provide VTuber's personal perspectives regarding avatarization, which can be compared or contrasted with the viewer perspective in this work. 

\section{Conclusion}

We explored interactions that viewers have with VTubers and the relationships that develop between them, focusing on avatarization. Semi-structured interviews with viewers outlined findings about how VTubers are performers that uniquely incorporate identities of both a real performer and a virtual avatar, positioning them as entertainers lying in a unique space between reality and virtuality. Viewers interact with these identities in distinct manners, and the fluid balance of identities impacts their motivations and relationships with VTubers. We also explored the novel interactions that are possible due to their increased virtuality --- avatarization affects both the more objective factors, such as the shape of the livestream, as well as introspective factors, such as viewer motivation and the audience-streamer relationship. We contextualized avatarization around prior livestreaming research and acknowledged how the unique position of VTubers generates a distinct set of limitations and challenges, to which we propose future areas of research. Ultimately, the popularity of VTubers represents a broader trend towards avatarized entertainers; we believe that the implications of this study are far-reaching in discussing general user experiences in consuming such new mediums of entertainment. 

\begin{acks}
This work was supported in part by the Natural Science and Engineering Research Council of Canada (NSERC) under Discovery Grant RGPIN-2019-05624.
\end{acks}

\bibliographystyle{ACM-Reference-Format}
\bibliography{sample-base}

\end{CJK*}
\end{document}